\documentclass[floatfix, preprintnumbers, twocolumn, amsmath, amssymb, 10pt]{revtex4}
\usepackage{graphicx}
\usepackage{dcolumn}
\usepackage{bm}
\usepackage{epsfig}

\begin{document}

\title{Radion Stabilization by Stringy Effects in General
Relativity}

\author{Subodh P. Patil$^{1)}$ and Robert Brandenberger$^{2,1}$}

\address{1) Department of Physics, Brown University, Providence, RI 02912,
USA\\
2) Physics Department, McGill University, 3600 rue Universit\'e,\\
Montreal, QC H3A 2T8, CANADA \\
E-mail: patil,rhb@het.brown.edu}

\begin{abstract}
We consider the effects of a gas of closed strings (treated quantum 
mechanically) on a background where one dimension is compactified on a 
circle. After we address the effects of a time dependent background on 
aspects of the string spectrum that concern us, we derive the energy-momentum 
tensor for a string gas and investigate the resulting space-time dynamics. We 
show that a variety of trajectories are possible for the radius of the 
compactified dimension, depending on the nature of the string gas, including 
a demonstration within the context of General Relativity (i.e. without a 
dilaton) of a solution where the radius of the extra dimension 
oscillates about the self-dual radius, without invoking matter that 
violates the various energy conditions. In particular, we find that in the case 
where the string 
gas is in thermal equilibrium, the radius of the compactified dimension 
dynamically stabilizes at the self-dual radius, after which a period of 
usual Friedmann-Robertson-Walker cosmology of the three uncompactified
dimensions can set in. We show that our radion stabilization mechanism requires a 
stringy realization of inflation as scalar field driven inflation invalidates our 
mechanism. We also show that our stabilization mechanism is consistent with 
observational bounds. 
\end{abstract}

\maketitle


\section{Introduction}

In the early days of string cosmology, it was realized that superstrings 
had an effect on space-time dynamics that was qualitatively quite different 
from that of particles or fields. In particular, it was realized that 
string winding modes could provide a confining mechanism for certain compact 
directions in such a way as to allow only three spatial dimensions to grow 
large \cite{BV}. Key to this realization are the T-duality of the
spectrum of string states, and the fact that the background is
described by Dilaton Gravity, and not by General Relativity with a
fixed dilaton (this is crucial in order that the background
equations obey the T-duality symmetry). The arguments of \cite{BV}
were put on a firmer basis by the analysis of \cite{TV} (see also
\cite{Ven91}). 

Starting point of the considerations of \cite{BV} is the assumption
that all spatial dimensions begin at close to the self-dual
radius (the string scale), and that matter consists of a hot gas
of string states. The considerations of \cite{BV}
were more recently applied to ``brane gas cosmology'' \cite{ABE,BEK},
a scenario in which the initial string gas is generalized to be a gas of all
brane modes. It was shown that given the hot dense initial conditions
assumed in \cite{BV}, the string winding modes are the last modes
to fall out of equilibrium and thus dominate the late time dynamics.
Hence \cite{ABE}, the inclusion of brane degrees of freedom does not
change the prediction that only three dimensions to grow large. 
The dynamical equations describing the growth of the three dimensions
which can become large were solved in \cite{BEK} (see also \cite{Campos1}).
In \cite{WB1}, it was shown that isotropy in these large dimensions is a 
consequence of the dynamics. In \cite{WB2}, it was found that if both
the momentum and winding modes of the strings are included in the
dynamical equations, the radius of the compactified dimensions 
is stabilized at the self-dual radius. More precisely, the expansion
of the three large dimensions leads to damped oscillations in the ``radion''
about the self-dual value. Thus, in the context of a background
described by Dilaton Gravity, radion stabilization is a natural
consequence of brane gas cosmology \footnote{See e.g. \cite{others} for
other recent papers on brane gas cosmology, and \cite{Kripfganz} for an
early work on the cosmology of string winding modes.}.

At the present time, however, the dilaton is most likely fixed (see, however,
\cite{Damour} for an alternate scenario). Thus, it is of interest to
explore how the inclusion of string (and brane) winding and momentum
modes influences the dynamical evolution of the radion in a background
space-time described by General Relativity (GR). There is another
motivation for studying this issue. Another corner of the M-theory
moduli space is 11-d supergravity. In \cite{Col2} it was found
that a brane gas in this background also admits a region in the phase
space of initial conditions in which only three spatial dimensions can
become large, although this corner may not be consistent with holographic
entropy bounds \cite{Col3} (see also \cite{Biswas} where the
considerations in this corner of M-theory moduli space was extended
to spaces with more general topologies). Motivated by these considerations,
we in this paper study a simplified problem, namely the questions
of how a gas of winding and momentum modes of strings winding one
compactified spatial dimension (taken to be a circle) 
effects the evolution of the radius (the radion). We start with
initial conditions in which the three spatially non-compact dimensions
are expanding. We find that the gas of string winding and momentum
modes gives a natural radion stabilization mechanism.
Our approach is to consider the effect of strings on 5D space-time 
dynamics (with the extra spatial dimension compactified to a circle)
by adding the appropriate matter term to the standard Einstein-Hilbert action. 
We will derive this term shortly
(see also \cite{Boehm} for a similar derivation). The resulting 
energy-momentum tensor leads to a novel behavior 
when inserted into the Einstein equations. We will find that we can generate 
a non-singular bouncing solution for the radius of the compactified 
dimension in the context of GR (without a dilaton) while respecting the 
Dominant Energy Condition for the matter content. Specifically, the
radion performs damped oscillations about the self-dual radius. 
Initially,
we study a pure state of matter with specific quantum numbers
obeying the T-duality symmetry. However, we will find that we can 
rather naturally extend the analysis to a gas of these strings in 
thermal equilibrium (with a bath of gravitons and photons), with the result 
that the radius of the compact dimension is dynamically stabilized at the 
self dual radius $R = \sqrt{\alpha'}$, where $2 \pi \alpha'$ is the 
inverse of the string tension (see also \cite{Borunda} for a
study of string gases in thermal equilibrium). 

In addition, we find that our model evolves according to standard 
Friedmann-Robertson-Walker (FRW) cosmology after the compact dimension has 
been stabilized, and that the resultant stabilization is incompatible with 
any subsequent inflationary epoch driven by a bulk scalar field
(for string-specific ideas on
how to generate inflation in brane gas cosmology see \cite{BEM}). 
However this conclusion can be avoided if some form of stringy inflation is 
realised where strings are produced in re-heating.

Before we can turn to any of this, we will have to address a 
question of principle concerning the string spectrum in a cosmological 
context (this issue is also being studied in \cite{Watson}). 
The question of formulating String Theory in a time-dependent
background is a current and active area of research. However, we are 
primarily interested in the behavior of strings in a background that 
evolves on a cosmological time scale. As can be seen from the FRW
equations, the cosmological time scale $H^{-1}$ (where $H$ is the
Hubble expansion rate) is larger than the characteristic microscopic 
time $\sigma^{-1}$ (where $\sigma^4$ is
the matter energy density) by a factor of $m_{pl} / \sigma$, where
$m_{pl}$ is the Planck mass. Thus, away from singular epochs in the 
history of the Universe, the cosmological time scale is going to be many, 
many orders of magnitude longer than the characteristic time scale of the 
string dynamics, and hence we should be able to 
inherit many of the features of the 
string spectrum in a static space-time (with some obvious modifications). 
We justify this intuition more rigorously in the Appendix, but we feel 
that it might suffice at this point to remind the reader of the approximate 
irrelevance of a time dependent background for a much more familiar theory: 
Quantum Field Theory (QFT). Although quantum fields in curved spaces exhibit 
several qualitatively different features from quantum fields in flat 
spaces\footnote{Examples are particle creation, non-uniqueness of the 
vacuum, non-trivial issues concerning existence of asymptotic states.}, 
we still manage to do a lot of sensible (and spectacularly successful) 
flat space-time QFT calculations 
despite the persistent Hubble expansion of space-time. The reason for this is 
easy to see: the contributions to masses, to scattering amplitudes,
to the structure of the Hilbert space of our theory, etc., that come 
from terms that depend on derivatives of the metric are in the present 
epoch highly suppressed and irrelevant. This is partly captured by the 
Adiabatic Theorem, which is the 
statement that given two systems with Hamiltonians that can be continuously 
interpolated, then in a precise sense, the eigenstates of the initial 
system will evolve into the eigenstates of the final system if this 
interpolation takes place slowly enough. Slow enough in simple quantum 
systems usually means that the variation happens over much longer time scales 
than the characteristic time of the system (by which we mean the time 
associated with the typical energy of the system:$\tau  \sim\frac{1}{E}$). 
Having said this, were we to study QFT in places where the metric varies a 
lot more rapidly (at the edges of black holes or in the very early Universe) 
we invariably have to account for the curvature of space. Thus, 
we can hope that the effects of a time dependent background on 
the closed string spectrum only require minor modifications to the flat 
space spectrum, if this time dependence is slow compared to the 
characteristic time of the string dynamics. We show in the Appendix that this 
is indeed the case, and in what follows we will stay within this regime.    

The Outline of this paper is as follows: we first derive the energy-momentum
tensor of a string gas (the derivation here is more general than the
one given in \cite{Boehm}). We
then insert this tensor into the Einstein equations and study the dynamics
of the radius of the compact dimension, assuming that the three large
spatial dimensions are in the expanding phase. First, we consider
a pure state of matter. Next, we extend the discussion to a thermal
state. In Section IV, we discuss
the late time dynamics and show that the stabilization of the radion is
not compatible with inflation in the three large spatial dimensions, assuming
the simplified description of matter which we are using. 

A few words on our notation: Greek indices typically stand for
5-dimensional space-time indices, Roman indices $i, j, ...$ are
associated with the non-compact spatial dimensions, and Roman
indices $a, b, ...$ are string world-sheet coordinates. 
The 5-dimensional Planck mass is denoted by $ M_{pl_5}$ (or $M_5$ in 
abbreviated form). We also work in natural units ($c = \hbar = k_B = 1$) 
where we pick energy to be measured in electron volts.

\section{The Energy-Momentum Tensor}

To study how a gas of strings affects space-time dynamics, we need to 
derive the energy-momentum tensor of such a gas. We begin by studying the 
energy-momentum tensor of a single closed string. Starting with the 
Nambu-Goto action
\begin{equation}
\label{NG}
S_{NG} = \frac{-1}{2\pi\alpha'}\int d^2 \sigma \sqrt{-h} \, ,
\end{equation}
where $h_{ab}$ denotes the world sheet metric
\begin{equation}
\label{ind}
h_{ab} = \partial_a X^\mu \partial_b X^\nu g_{\mu \nu}(X) 
\end{equation}
(and $h$ is its determinant),
we see that any variation in the space-time metric 
$g_{\mu \nu}$ induces a variation in
the induced world-sheet metric (where the unmatched indices
indicate that we perturb only the $\lambda \beta$ component of the metric):
\begin{eqnarray*}
g_{\mu \nu}(X) &\rightarrow& g_{\mu \nu}(X) 
+ \underbrace{\delta^\lambda_\mu \delta^\beta_\nu \delta^D(X^\tau- y^\tau)}_{\delta g_{\mu \nu}}\\ 
h_{ab}(\sigma) &\rightarrow& h_{ab} 
+ \underbrace{\partial_a X^\lambda \partial_b X^\beta \delta^D(X^\tau-y^\tau)}_{\delta h_{ab}}
\end{eqnarray*}

Now, varying the Nambu-Goto action \begin{it}with respect to the space-time 
metric \end{it} (performing a perturbation $\delta g_{\alpha \beta}$
which acts on the metric as given above) will give us the 
space-time energy-momentum tensor of a single string:
\begin{eqnarray*}
\frac{\delta S_{NG}}{\delta g_{\lambda\beta}(y)} &=& -\frac{1}{4\pi\alpha'} 
\int d^2 \sigma \sqrt{-h}h^{ab}\delta h_{ab}\\ 
&=& -\frac{1}{4\pi\alpha'} \int d^2 \sigma \sqrt{-h}h^{ab}
\partial_a X^\lambda \partial_b X^\beta \delta^D(X^\tau-y^\tau) \, .
\end{eqnarray*}
We must first discuss the meaning of the expression
\begin{eqnarray} 
\int d^2 &\sigma& \delta^D(X^\tau-y^\tau) \nonumber \\ 
&=&
\int d^2 \sigma \delta(X^0-y^0)\delta(X^1-y^1)...\delta(X^D-y^D)
 \, . \nonumber
\end{eqnarray}
In order to change the variable of integration, we need to apply 
$d\sigma^a = \frac{dX^\lambda}{\partial_a X^\lambda}$ 
and sum over all the zeroes of 
$X^\lambda [\sigma] - y^\lambda$ when performing the integration. 
However, since we considering modes winding one particular spatial
direction, there are precisely two coordinates that are monotonic 
functions of the world-sheet parameters: $X^0$ being a monotonic function 
of $\sigma^0$ and $X^D$ being a monotonic function of $\sigma^1$ 
(the $D^{th}$ direction is taken to be compact). Thus,
\begin{eqnarray*}
&d^2& \sigma \delta^D(X^\tau[\sigma]-y^\tau)\\ 
&=&  d\sigma^0 d\sigma^1 \delta(X^0[\sigma] -y^0)\delta(X^1[\sigma]-y^1)...
\delta(X^D[\sigma]-y^D)\\ 
&=&
\frac{dX^0}{|{\stackrel{.}{X^0}}|}\frac{dX^D}{|X'^D|}\sqrt{-g_{00}g_{DD}}
\\ && \times \delta(X^0-y^0)\delta(X^D-y^D)\delta^{D-2}(X^i-y^i) \, ,
\end{eqnarray*}
where we include the metric factors in the last line so that we can take 
the delta functions in $X^0$ and $X^D$ to be properly normalized. With 
this result, we get:
\begin{eqnarray*}
&& \frac{\delta S_{NG}}{\delta g_{\lambda\beta}} 
= -\frac{\delta^{D-2}(X^i-y^i)}{4\pi\alpha' \sqrt{-g^{00}g^{DD}}}\\ 
&&\times 
\int \frac{dX^0}{|\stackrel{.}{X^0}|}\frac{dX^D}{|X'^D|} \delta(X^0-y^0)\delta(X^D-y^D)\sqrt{-h}h^{ab}\partial_a X^\lambda \partial_b X^\beta\\ 
&& = -\frac{1}{4\pi\alpha' \sqrt{-g^{00}g^{dd}}} \\ && \times
\frac{\delta^{D-2}(X^i-y^i)}{|\stackrel{.}{X^0}X'^D|}\sqrt{-h}h^{ab}\partial_a X^\lambda \partial_b X^\beta \Big|_{X^0=y^0,X^D=y^D} \, ,
\end{eqnarray*}
where we use the inverse metric to write the metric contributions in the 
denominator. Thus, the single string space-time energy-momentum tensor
becomes
\begin{eqnarray}
\label{em}
T^{\lambda \beta} 
&=& \frac{-2}{\sqrt{-g}}\frac{\delta S}{\delta g_{\lambda\beta}} \\
&=& \frac{1}{2\pi\alpha'} 
\frac{\delta^{D-2}(X^i-y^i)}{|\stackrel{.}{X^0}X'^D| \sqrt{-g}}
\frac{\sqrt{-h}h^{ab}\partial_a X^\lambda 
\partial_b X^{\beta}}{\sqrt{-g^{00}g^{DD}}} \, . \nonumber
\end{eqnarray}
Inserting the explicit form of the inverse world-sheet metric
\begin{equation}  
\label{inverse}
h^{ab} = \frac{1}{h} \begin{pmatrix} h_{22} & -h_{12} \cr -h_{21} & h_{11}\end{pmatrix} = 
\frac{1}{h} 
\begin{pmatrix} X'^\mu X'_\mu & -\stackrel{.}{X^\mu}X'_{\mu} \cr -\stackrel{.}{X^\mu}X'_{\mu} & \stackrel{.}{X^\mu}\stackrel{.}{X_\mu}
\end{pmatrix}
\end{equation}
and using the constraints on the world-sheet fields
\footnote{Which come from working in conformal gauge $h_{ab} = diag(-1,1)$}
\begin{equation}
\label{const1}
P_\mu X'^\mu = 0 
\end{equation}
\begin{equation}
\label{const2}
P_\mu P^\mu + X'_\mu X'^\mu = 0 \, ,
\end{equation}
we can write (\ref{em}) as
\begin{equation}
\label{em2}
T^{\lambda \beta} = \frac{-1}{2\pi\alpha'} \frac{\delta^{D-2}(X^i-y^i)}{|\stackrel{.}{X^0}X'^D|\sqrt{-det g_{ij}}}[ X'^\lambda X'^\beta - \stackrel{.}{X}^\lambda \stackrel{.}{X}^\beta ]
\end{equation}

Next, we solve for $\stackrel{.}{X^0}$ using the constraint (\ref{const2})
which becomes
\begin{eqnarray*}
0 &=& -\stackrel{.}{X^0}\stackrel{.}{X^0} + \stackrel{.}{X^i}\stackrel{.}{X_i} + \stackrel{.}{X^D}\stackrel{.}{X^D} \\ 
&& - X'^0 X'^0 + X'^i X'_i + X'^D X'_D \, ,
\end{eqnarray*}
where we have explicitly used the background metric
\begin{equation}
\label{metric0}
g_{\mu \nu} = diag(-1,a^2(t),a^2(t),a^2(t),b^2(t)) \, .
\end{equation}
It is consistent with the equations of motion in a (slow enough) time 
varying background to set $X'^0 = 0$ \footnote{See the Appendix for all 
statements made in this section concerning results that are valid in a 
time dependent background}, so that
\begin{equation}
\stackrel{.}{X^0}^2 = P^iP_i + X'^iX'_i + P^D P_D + X'^D X'_D \, . 
\end{equation} 
In addition (in a slowly time dependent background) the right hand side can 
be expressed in terms of the familiar oscillator expansion. Accounting for 
the zero mode operators explicitly, we get the center of mass momentum 
from the spatial zero modes and the winding energy from the zero mode 
terms in the compactified direction. The other modes give us the
left and right moving oscillator terms (see \cite{Pol} for details):
\begin{equation}
\label{x0}
\stackrel{.}{X^0} = \sqrt{ g^{ij}p_i p_j + 
\frac{2}{\alpha'}(N + \stackrel{\_}{N} - 2) + (\frac{n}{b})^2 + (\frac{wb}{\alpha'})^2} \, ,
\end{equation} 
where $n$ and $w$ are the quantum numbers for momentum and winding
in the compact direction, respectively, and $N$ and $\stackrel{\_}{N}$ are the
levels of the left- and right-moving oscillator modes of the string,
respectively. 
The expression above is none other than the energy of the string. 
Using the level matching constraints 
\begin{equation} \label{match}
N + nw - \stackrel{\_}{N} = 0 \, ,
\end{equation}
we finally end up with \footnote{See the second footnote in the Appendix
where we 
remind the reader why the momenta must be contracted with the inverse metric.}
\begin{equation}
\label{x0E}
\stackrel{.}{X^0} = \sqrt{ g^{ij}p_i p_j + \frac{4}{\alpha'}(N - 1) + (\frac{n}{b} + \frac{wb}{\alpha'})^2} \, .
\end{equation}  

Now, we are ready to evaluate (\ref{em2}) for a single wound string. 
We have an explicit expression for $\stackrel{.}{X^0}$ and we know 
that $|X'^5| = |w|b$ in units of $\alpha'$ for a wound string, 
the factor of $|w|$ being canceled by the summation over all 
($w$ in total) zeroes of the argument of the delta function. 
Thus, component by component, we get:
\begin{eqnarray}
\label{rho}
&&T^0_0 = -\rho \\
&&= -\frac{1}{2 \pi} \frac{\delta^{3}(X^i - y^i)}{a^3 b} 
\sqrt{p^i p_i + \frac{4}{\alpha'}(N - 1) + (\frac{n}{b} + \frac{wb}{\alpha'})^2} \nonumber
\end{eqnarray}      
\begin{eqnarray}
\label{p}
&&T^i_i = p \\
&&= \frac{1}{2\pi }\frac{\delta^{3}(X^i - y^i)}{a^3 b} \frac{p^ip_i}{\sqrt{ p^i p_i + \frac{4}{\alpha'}(N - 1) + (\frac{n}{b} + \frac{wb}{\alpha'})^2}}
\nonumber
\end{eqnarray}
\begin{eqnarray}
\label{r}
&&T^5_5 = r \\
&&= \frac{1}{2\pi }\frac{\delta^{3}(X^i - y^i)}{a^3 b}\frac{ \frac{n^2}{b^2} - \frac{w^2 b^2}{\alpha'^2} }{\sqrt{ p^i p_i + \frac{4}{\alpha'}(N - 1) + (\frac{n}{b} + \frac{wb}{\alpha'})^2}} \nonumber
\end{eqnarray}
(note that we label the extra spatial coordinate by ``5'')
where we ignore off-diagonal components since we are about to apply these
expressions to an isotropic gas of strings. 

However, we wish to present at 
this point another derivation of this result which is rather more direct. 
Consider the energy of a single wound string:
\begin{equation}
\label{en}
E^2 = p^i p_i + \frac{4}{\alpha'}(N - 1) + (\frac{n}{b} + \frac{wb}{\alpha'})^2 
\, .
\end{equation}
A spatially uniform gas of such strings with the same quantum numbers would 
have a 5-dimensional energy density
\begin{equation}
\label{end}
\epsilon = \frac{\mu(t)}{2\pi b}\sqrt{ p^i p_i + \frac{4}{\alpha'}(N - 1) + (\frac{n}{b} + \frac{wb} {\alpha'})^2} \, ,
\end{equation}
where $\mu(t)$ is the number density of strings. We divide by $2\pi b$ 
since this energy will be uniformly distributed over the length of the 
string. The momentum that appears in this expression is now the momentum 
squared of a gas of strings whose momenta have identical magnitudes, but 
whose directions are distributed isotropically. To fully account for the 
metric factors in this expression, we write $\mu(t)$ as $\mu_0(t)/a^3(t)$ 
since this is how a number density explicitly depends on the metric. 
Now, realizing that this is an energy density, we can introduce this gas of 
strings as matter interacting with the gravitational field by just adding 
the following term to the gravitational part of the action:
\begin{equation}
S_{int} = -\int d^5x \sqrt{-g} \epsilon  
\end{equation}
(see e.g. Section 10.2 of \cite{MFB}).

Realizing now that the metric factors in the denominator of the 
expression for the energy density can be written as 
$a^3 = \sqrt{det(g_{ij})}$ and $b = \sqrt{g_{55}}$, we can write the above 
equation as:
\begin{eqnarray*}
S_{int} &=& -\int \frac{d^5x \sqrt{-g}}{\sqrt{detg_{ij}} \sqrt{g_{55}}} 
\frac{\mu_0(t)}{2\pi} \\
&& \times \sqrt{p^i p_i + \frac{4}{\alpha'}(N - 1) + (\frac{n}{b} + \frac{wb}{\alpha'})^2} \\
&=& -\int d^5x \sqrt{-g_{00}} \frac{\mu_0(t)}{2\pi} \\ 
&& \times \sqrt{p^i p_i + \frac{4}{\alpha'}(N - 1) + (\frac{n}{b} + \frac{wb}{\alpha'})^2}
\end{eqnarray*}
By our metric ansatz and the isotropy of the distribution of the momenta, we 
have that $p^ip_i = a^{-2}(\frac{p^2}{3} + \frac{p^2}{3} + \frac{p^2}{3})$. 
Using this fact, it is straightforward to show that the energy-momentum 
tensor derived from this interaction term is:
\begin{equation}
\label{rho2}
T^0_0 = -\rho = -\frac{1}{2 \pi} \frac{\mu_0}{a^3 b} 
\sqrt{p^i p_i + \frac{4}{\alpha'}(N - 1) + (\frac{n}{b} + \frac{wb}{\alpha'})^2}
\end{equation}      

\begin{equation}
\label{p2}
T^i_i = p = \frac{1}{2\pi }\frac{\mu_0}{a^3 b} \frac{p^2/3}{\sqrt{ p^i p_i + \frac{4}{\alpha'}(N - 1) + (\frac{n}{b} + \frac{wb}{\alpha'})^2}}
\end{equation}

\begin{equation}
\label{r2}
T^5_5 = r = \frac{1}{2\pi }\frac{\mu_0}{a^3 b}\frac{ \frac{n^2}{b^2} - \frac{w^2 b^2}{\alpha'^2} }{\sqrt{ p^i p_i + \frac{4}{\alpha'}(N - 1) + (\frac{n}{b} + \frac{wb}{\alpha'})^2}}
\end{equation}

\noindent which is exactly what we would get from (\ref{rho}), (\ref{p})
and (\ref{r}) were we to construct a hydrodynamical average with an 
isotropic momentum distribution. 

We now investigate some 
simple aspects of our result. The first thing to note is that $T^5_5$, 
which is the pressure along the compact direction, gets a negative 
contribution from the winding of our strings and a positive contribution 
from the momentum along this direction. The spatial pressure is always 
positive, and for the simple case $n=w=0$, $N=1$, which describes a gas of 
gravitons moving in the non compact directions, we obtain 
$r = 0, p = \rho/3$. 

Since we are about to study the effects of this 
energy-momentum tensor on space-time, we should make sure that the
energy-momentum tensor is covariantly conserved, or else it will not be 
consistent to equate it to the covariantly conserved Einstein tensor.
The covariant conservation of $T^\mu_\nu$
\begin{equation}
0 = \nabla_\mu T^\mu_\nu \, , \nonumber
\end{equation}
where $\nabla_\mu$ is the covariant derivative operator, implies
\begin{eqnarray*}
0 &=& \stackrel{.}{\rho} + 3\frac{\stackrel{.}{a}}{a}(\rho + p) 
+ \frac{\stackrel{.}{b}}{b}(\rho + r)\\ 
0 &=& \partial_i p \\ 
0 &=& \partial_5 r \, .
\end{eqnarray*} 
It is straightforward to check that our energy-momentum tensor satisfies 
this as an identity. In the continuity equation, this is due to the metric 
factors contained in the energy density, which upon differentiation produce 
terms that exactly cancel the terms proportional to the Hubble factors. 
The remaining equations are trivially satisfied by our setup, which assumed 
an axis of symmetry along the compactified dimension (the Kaluza-Klein setup) 
with homogeneous and isotropic spatial sections.

One final point to note is that we have derived an energy-momentum 
tensor that exhibits positive pressures along the non-compact directions 
and positive or negative pressures along the compactified direction. 
We need to ensure that this negative pressure has a bounded equation of 
state as otherwise our theory would be unstable. The Dominant Energy 
Condition (DEC) of General Relativity \cite{DEC}
ensures the stability of the vacuum, and requires that the equation of 
state parameter $\omega = p / \rho$ be greater than or equal to -1 
(see e.g. \cite{Trodden} for a recent discussion). 
Since the spatial pressures are always positive, we only need to check 
our equation of state for the pressure along the compact direction:
\begin{eqnarray*}
r &=& \frac{1}{2\pi}\frac{\mu_0}{a^3 b}\frac{ \frac{n^2}{b^2} - \frac{w^2 b^2}{\alpha'^2} }{\sqrt{ p^i p_i + \frac{4}{\alpha'}(N - 1) + (\frac{n}{b} + \frac{wb}{\alpha'})^2}}\\
&=& \rho \times \frac{ \frac{n^2}{b^2} - \frac{w^2 b^2}{\alpha'^2}}{p^i p_i + \frac{4}{\alpha'}(N - 1) + (\frac{n}{b} + \frac{wb}{\alpha'})^2} \, ,
\end{eqnarray*}
where the co-efficient of $\rho$ in the above is our equation of state parameter. 
Were we to consider states described by $n = \pm 1$, $w = -n$, $N = 1$ (which as 
we will see further on, turn out to be the relevant states that give us 
stabilization), this parameter remains bounded as $b$ varies
\footnote{Where key to this is the observation that as we approach the 
value $b = \sqrt{\alpha'}$, these states become massless, and acquire a 
non-zero momentum along the non-compact directions, which depends on the 
ambient temperature. If this momentum is large enough, we can be assured 
that (\ref{bdd}) is satisfied.}:

\begin{equation}
\label{bdd}
-1 \leq \omega \leq 1
\end{equation}

Thus, we have
verified that the spectrum of string states satisfies the DEC, and in 
doing so ensured ourselves of sensible space-time dynamics arising from 
the string gas, the topic we will turn our attention to next.     

\section{Space-Time Dynamics}

We start with the Einstein tensor derived from the metric (\ref{metric0}):
\newcommand{\tdot}[1]{\stackrel{.}{{#1}}}
\newcommand{\hub}[1]{\frac{\stackrel{.}{{#1}}}{{#1}}}
\newcommand{\tddot}[1]{\stackrel{..}{{#1}}}
\newcommand{\hubd}[1]{\frac{\stackrel{..}{{#1}}}{{#1}}}
\begin{eqnarray*}
G^0_0 &=& -3\hub{a} \Bigl[ \hub{a} + \hub{b} \Bigr] \\ G^i_j &=& -\delta^i_j \Bigl[ 2\hubd{a} + \hubd{b} + \Bigl( \hub{a} \Bigr)^2 +2\hub{b}\hub{a} \Bigr]\\ G^5_5 &=& -3 \Bigl[ \hubd{a} + \Bigl( \hub{a} \Bigr)^2 \Bigr] \, .\\
\end{eqnarray*}
Equating this to $\frac{1}{M^3_{pl_5}} T^\mu_\nu$ will give us the Einstein 
equations. However, let us focus on the equation that governs the evolution 
of the scale factor $b$. Starting with $G^i_j$ and eliminating 
$\tddot{a}$ and $\tdot{a}^2$ by adding the appropriate combinations of 
$G^0_0$ and $G^5_5$, we get:
\begin{equation}
\label{beom}
\tddot{b} + 3H\tdot{b} + \frac{b}{M^3_{pl_5}}\Bigl(p - \frac{2r}{3} - \frac{\rho}{3} \Bigr) = 0 \, ,
\end{equation}
where $H$ is the 3-dimensional Hubble factor. This is a second order, 
nonlinear (because of the $b$ dependence in the matter terms) differential 
equation with a damping term and a driving term. We will demonstrate further 
on that the Einstein equations admit expanding solutions for the non-compact 
dimensions $(H>0)$, and take it as a given for what follows. Thus, in spite of 
its non-linearity, we easily see that (\ref{beom}) describes an expanding or 
a contracting scale factor depending on the sign of the driving term. 

The first thing to notice from this equation is that matter for which the 
quantity $p - \frac{2r}{3} - \frac{\rho}{3}$ vanishes will not contribute to 
the dynamics of the compact dimension. Thus, recalling that a gas of 
gravitons ($n=w=0$, $N=1$) has an equation of state $p = \frac{\rho}{3}$, 
$r = 0$, as does a gas of ordinary 4-dimensional photons, we see that such 
matter will not affect the dynamics of the scale factor $b$. In fact, such
a background 4-dimensional gas provides an excellent candidate for a
thermal bath which we will eventually couple our gas of winding modes to. 

First,
however, we will study this driving term as it is, for a gas  consisting of
strings with identical quantum numbers. Upon evaluating the driving term, 
we find that:
\begin{eqnarray}
\label{drive}
\frac{b}{M^3_{pl_5}}\Bigl(&p& - \frac{2r}{3} - \frac{\rho}{3} \Bigr) 
= \frac{\mu_0}{M^3_{pl_5} a^3 2\pi } \\ & \times &
\frac{ -\frac{n^2}{b^2} - \frac{2nw}{3 \alpha'} + \frac{w^2b^2}{3\alpha'^2} - \frac{4(N-1)}{3\alpha'}}{\sqrt{ p^i p_i + \frac{4}{\alpha'}(N - 1) + (\frac{n}{b} + \frac{wb}{\alpha'})^2} } \, , \nonumber
\end{eqnarray}
from which we infer that momentum modes and oscillator modes lead to  
expansion of the scale factor, whereas the winding modes produce contraction. 
Exactly what happens, of course, depends on the values of the quantum numbers. 
It should be recalled that the quantum numbers are subject to the constraint 
$nw + N \geq 0$ coming from the level matching conditions 
(see Eq. \ref{match}). 

Let us pick a particular set of 
quantum numbers. As we shall see later, the most interesting case is 
when $n=-w = \pm 1$, $N=1$, in which case the driving term becomes:
\newcommand{\twid}[1]{\stackrel{\_}{{#1}}}
\begin{equation} \label{pot}
\frac{b}{M^3_{pl_5}}\Bigl(p - \frac{2r}{3} - \frac{\rho}{3} \Bigr) = \frac{2 \mu_0}{M^3_{pl_5} a^3 2\pi \sqrt{\alpha'}}\frac{ -\frac{1}{\twid{b}^2} + \frac{2}{3} + \frac{\twid{b}^2}{3} }{\sqrt{ \alpha' p^i p_i + (\frac{1}{\twid{b}} - \twid{b})^2} } \, ,
\end{equation}
where $\twid{b}$ is the scale factor in units of $\sqrt{\alpha'}$. 
Quite generically, we can explore features of the ``potential" energy that 
will yield this driving term. We see that, since the denominator is strictly 
positive, and the driving term changes sign at $\twid{b} = 1$, this value
will be a minimum of the potential energy, and hence a point of equilibrium. 
Numerical integration of the driving term yields the potential energy 
curve of Figure 1 \footnote{We model the momentum squared as a smooth 
function of 
$\twid{b}$ such that it takes on some non zero value at $\twid b = 1$ and 
falls off on either side. This is because at $\twid{b} = 1$ the state 
described by $n=-w=1$, $N=1$ becomes massless and should have a finite 
non zero momentum, but as the scale factor increases or decreases the state 
becomes more and more massive and hence the momentum becomes negligible in 
comparison. The generic feature of a minimum at $\twid{b} = 1$ is robust, 
however, since as we mentioned earlier, the force term will always change 
sign at $\twid{b} = 1$. One finds that the minimum is always fairly concave 
independent of the nature of the $b$ dependence of the momentum.} 
where the potential is plotted in units of 
$\frac{2 \mu_0}{M^3_{pl_5} a^3 2\pi \sqrt{\alpha'}} $ 
as a function of $\twid{b}$.

\begin{figure}
\epsfxsize=2.9 in \epsfbox{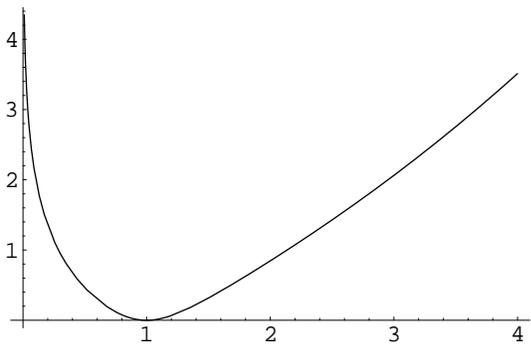}
\caption{Potential term for $n=-w= \pm1, N=1$. The horizontal axis is $b$ (in
string units), the vertical axis gives the potential in units of ${\cal U}$,
where ${\cal U}$ is the prefactor on the right hand side of (\ref{pot}).}
\end{figure}

Because of the Hubble damping term in the equation of motion for $\twid{b}$ 
(which is obtained by dividing (\ref{beom}) through by a factor of 
$\sqrt{\alpha'}$), the scale factor will perform damped oscillations about
the minimum of the potential to which it will evolve with rapidity depending 
on the value of the ``spring constant" multiplying the driving term:
\begin{equation}
\label{spring}
k = \frac{2 \mu_0}{M^3_{pl_5} a^3 2\pi \alpha'} \, .
\end{equation} 
Thus, we have established that a gas of string modes with non zero winding and
momentum numbers in the compact direction will provide a dynamical
stabilization mechanism for the radion, provided that the three non-compact
dimensions are expanding (such a behavior was already found in an early
study \cite{Kripfganz} of the dynamics of string winding modes - we
thank Scott Watson for drawing our attention to this paper). We will 
address the phenomenology of this stabilization mechanism further on, 
simply stating for now that we can obtain a robust stabilization mechanism 
which is consistent with observational bounds.

At this point, we wish to mention that the ``Quantum Gravity Effects"
required 
to stabilize the extra dimensions in earlier attempts \cite{KK1,KK2} 
at Kaluza-Klein cosmology find a stringy realization here, in that all that 
was required for radius stabilization was matter that depended on the 
size of the extra spatial dimensions in a non-trivial way. 

To round off the discussion, we 
wish to demonstrate that our assumption of an expanding scale factor $a(t)$ 
is consistent with an oscillating scale factor $b(t)$. Consider the two 
Einstein equations that do not contain second derivatives of $b$:
\begin{eqnarray*}
\frac{\rho}{3M^3_5} &=& H^2 + H\hub{b} \, ,\\ 
\frac{-r}{3M^3_5} &=& \tdot{H} + 2H^2 \, .
\end{eqnarray*}
These equations imply that
\begin{equation}
\tdot{H} - 2H\hub{b} = -\frac{1}{3M^3_5}(2\rho + r) \, .
\end{equation}
The resulting equation for $H$ has the integrating factor $1/b^2$, and 
hence the solution:
\begin{equation}
\label{hubtime}
H(t) = H_0\Bigl(\frac{b(t)}{b_0}\Bigr)^2 - \frac{b^2(t)}{3M^3_5}\int_0^t dt' \frac{(2\rho + r)}{b^2(t')} \, .
\end{equation}
Now, from the discussion surrounding (\ref{bdd}), we see that 
$\rho \geq r \geq -\rho$. Thus, the contribution to the integral in the
above is strictly positive, and the second term on the right hand side
of (\ref{hubtime}) can at most take on the value:
\begin{equation}
\label{hubpos}
\frac{b^2(t)}{M^3_5}\int_0^t dt' \frac{\rho(t')}{b^2(t')} \, .
\end{equation} 
Thus, we see that if we pick the initial conditions for $H$ appropriately, 
the scale factor $a$ can be taken to be expanding $(H > 0)$ regardless of 
the detailed motion of $b$. In fact, if we assume that $H_0$ starts out 
positive, then $H(t)$ will remain so if
\begin{equation}
\label{bounce}
H_0 \geq \frac{1}{3M^3_5}\int_0^t dt'\frac{2\rho + r}{b^2(t')/b^2_0} \, ,
\end{equation}
where the eventual stabilization of $b$ and the $1/Vol$ dependence of $\rho$ 
and $r$ will bound the integral, which implicitly depends on 
$H$ itself. This implicit dependence works in our favor in that the 
larger we make $H_0$, the smaller the integral becomes and so we can 
imagine picking an initial $H_0$ such that a persistent expansion of the 
non-compact dimensions results. Note, however that if in the spirit
of brane gas cosmology, we assume that all spatial dimensions are starting
out with the same size and instantaneously static, then it may not be
possible to evolve to a situation in which three large spatial dimensions
are expanding. This is, in fact, the result that emerges from the work of
\cite{Col3}, at least in a certain region of phase space. 

\section{Thermal String Gases}

In what we have done so far, we have just considered the behavior of the 
size of the extra dimension in a rather artificial setting, namely imposing 
a gas of strings with a fixed set of quantum numbers. One expects the early 
Universe to be in a state of thermal agitation, and it is inevitable that 
transitions between different energy levels will be induced in the string gas. 
Thus, to have any hope of realistically applying our setup to cosmology, 
we need to study the effects of placing the string gas in a thermal bath. 
Referring to our expression for the energy density of a string (\ref{rho2}), 
we see that a gas of strings with different quantum numbers will have the 
energy density:
\begin{equation}
\label{rhotherm}
\rho = \sum_{n,w,N,p^2} \frac{\mu_{n,w,N,p^2}}{a^3b2\pi } \sqrt{ p^i p_i + \frac{4}{\alpha'}(N - 1) + (\frac{n}{b} + \frac{wb}{\alpha'})^2}
\end{equation}
with densities $\mu_{n,w,N,p^2}$ for each given set of quantum numbers. The
expressions for the pressure terms $p$ and $r$ are similarly modified.
If we are in thermal equilibrium, the densities are given
by the Boltzmann weight
\begin{equation}
\label{boltz}
\mu_{n,w,N,p^2} = e^{\beta E_{ref}}e^{-\beta E_{n,w,N,p^2}}\mu_{ref} \, ,
\end{equation}
where the subscript ``ref" refers to some arbitrary reference energy
level. 

What constitutes the thermal bath to which the string gas is coupled to? 
We know from the discussion at the end of Section II that gravitons described 
by unwound strings propagate in the non-compact directions with an equation 
of state $p = \rho/3$. Introducing a gas of ordinary photons will also 
add a 4-dimensional component to the energy-momentum 
tensor with the same equation of state. Such particles offer us an 
ideal candidate for a thermal bath, for two reasons. Firstly, thermal 
equilibrium demands a coupling of some kind between the gas of winding modes 
and the gas of gravitons and photons. Such a coupling is readily provided by 
the tree-level reaction $w + \stackrel{\_}{w} \to h_{\mu \nu}$ via which 
winding modes of equal and opposite winding scatter to produce 4-d gravitons. 
This thermalization mechanism will, at non-zero temperatures, create an 
equilibrium where there will be an ever-present non-zero winding mode density 
due to gravitons scattering into winding modes (and vice-versa). This thermal 
bath has the further property that it does not affect the dynamics 
of the extra dimension other than through the Hubble factor (which it 
influences), since the driving term is only sensitive to the combination 
$p - \rho/3 - 2r/3$ which vanishes for the graviton and photon components of 
the energy-momentum tensor. 

With the above in mind, the driving term for the 
scale factor $b$ becomes:
\begin{eqnarray}
\label{thermaldrive}
\frac{b}{M^3_{pl_5}}&&\Bigl(p - \frac{2r}{3} - \frac{\rho}{3} \Bigr) 
= \frac{\mu_{ref} e^{\beta E_{ref}}}{M^3_{pl_5} a^3 2\pi} \\ & \times &
\sum_{n,w,N,p^2} e^{-\beta E} \frac{ -\frac{n^2}{b^2} - \frac{2nw}{3 \alpha'} + \frac{w^2b^2}{3\alpha'^2} - \frac{4(N-1)}{3\alpha'}}{\sqrt{ p^i p_i + \frac{4}{\alpha'}(N - 1) + (\frac{n}{b} + \frac{wb}{\alpha'})^2} } \nonumber 
\, ,
\end{eqnarray}
where the Boltzmann weight in the summation depends on the values of the 
quantum numbers. We remind the reader that the sum is restricted by the level 
matching condition $N + nw \geq 0$. For completeness we also remind the 
reader of the resulting equation of motion for $b$:
\begin{eqnarray}
\label{beomtherm}
0 &=& \tddot{b} + 3H\tdot{b} + 
\frac{\mu_{ref} e^{\beta E_{ref}}}{M^3_{pl_5} a^3 2\pi} \\ && \times
\sum_{n,w,N,p^2} \frac{e^{-\beta E}}{ \sqrt{E}} \Bigl( -\frac{n^2}{b^2} - \frac{2nw}{3\alpha'} + \frac{w^2b^2}{3\alpha'^2} - \frac{4(N-1)}{3\alpha'} \Bigr)
\nonumber
\end{eqnarray}

The summation which has to be performed in order to obtain the driving term is 
quite formidable, were it not for a rather special feature of string 
thermodynamics. Consider the argument of the exponential in the Boltzmann 
factor:
\begin{eqnarray*}
\beta E_{n,w,N,p^2} &=& \beta\sqrt{ p^i p_i + \frac{4}{\alpha'}(N - 1) + (\frac{n}{b} + \frac{wb}{\alpha'})^2}\\&=& \frac{\beta}{\sqrt{\alpha'}} \sqrt{ \alpha' p^i p_i + 4(N - 1) + (\frac{n}{\twid{b}} + \twid{b}w)^2}
\end{eqnarray*}
We see that when the energy is expressed in terms of dimensionless variables, 
we pull out a factor of $\sqrt{\alpha'}$. Thus, the argument of the 
exponential in the Boltzmann weight is $\beta/\sqrt{\alpha'}$ times a term of 
order unity. To be able to neglect all but the first few terms in the 
summation, we need the Boltzmann factor to be considerably less than unity, 
i.e. that
\begin{equation} \nonumber
e^{-\frac{\beta}{\sqrt{\alpha'}}} \ll 1 
\, .
\end{equation}
Thus, if this condition is satisfied, then the terms which dominate
the sum will be those whose quantum numbers render them nearly massless, 
since any state with even one of its quantum numbers being different from the 
nearly massless combination will produce a term of order unity times 
$\beta/\sqrt{\alpha'}$. 

Let us take a closer look at the above 
condition. We know from string thermodynamics that there exists a limiting 
temperature -- the Hagedorn temperature 
$T_H$ \cite{Hage}. Thus, for us to even be able to apply 
thermodynamics, we need to be well below this temperature, which for all 
the string theories is of the order of $T_H \sim 1/\sqrt{\alpha'}$. Thus, 
$\beta_H \sim \sqrt{\alpha'}$, and so if we are at a temperature 
much lower than the Hagedorn temperature, i.e.
$T \ll T_H$ or equivalently $\beta_H \ll \beta$, then 
\begin{equation}
\sqrt{\alpha'} \ll \beta \, ,
\end{equation}    
which is exactly what we need for the Boltzmann weights of higher mass states
to be negligible.
So, even if the thermal bath has a temperature of only one order of magnitude 
below the Hagedorn temperature, then 
$e^{-\frac{\beta}{\sqrt{\alpha'}}} \sim 10^{-5}$ which clearly lets us ignore 
any term whose energy in dimensionless units 
$\sqrt{ \alpha' p^i p_i + 4(N - 1) + (\frac{n}{\twid{b}} + \twid{b}w)^2}$ 
is anything other than zero. This translates into us being able to 
\begin{it} neglect all states other than those that are massless.\end{it} 
Thus, the summation now becomes very tractable, 
and we can also have faith in our truncation of the string spectrum to the 
lightest states all the way up to very high temperatures $(T \sim T_H/10$). 
Before we carry on we should remark that exactly massless states have a 
non-zero momentum given by the thermal expectation value of 
$E = |p| = 3/\beta$.

Let us then proceed to evaluate (\ref{thermaldrive}), so that we can evolve 
$b$ in time using (\ref{beomtherm}), recalling that now we only sum over the 
massless and near massless states subject to the level matching constraint. 
Let us begin near $\twid{b} = 1$, i.e. $\twid{b} = 1 + \Gamma$. Then for the 
case that $\Gamma \neq 1$, we only have one truly massless state: 
$n=w=0$, $N = 1$. This term will not contribute to the driving force for
$b$ since
\begin{equation}
\frac{\Bigl( -\frac{n^2}{b^2} - \frac{2nw}{3\alpha'} + \frac{w^2b^2}{3\alpha'^2} - \frac{4(N-1)}{3\alpha'} \Bigr)}{|p|} = 0 \, .
\end{equation}
Thus, the next lightest state which has quantum numbers $N=1, n=-w = \pm1$
will dominate the evolution of $b$. The level matching constraints 
$N + nw \geq 0$ ensure that there are no more nearly massless states 
(Note we only consider states with positive mass squared - any tachyonic 
states are posited to be absent from our spectrum). Such states will 
contribute:
\begin{equation} \label{source}
\frac{e^{\frac{-\beta}{\sqrt{\alpha'}}\sqrt{(\frac{1}{\twid{b}} - \twid{b})^2 + \alpha'p^2}}\Bigl(-\frac{1}{\twid{b}^2} + \frac{\twid{b}^2}{3} + \frac{2}{3}\Bigr)}{\sqrt{( \frac{1}{\twid{b}} - \twid{b})^2 + \alpha'p^2}} \, .
\end{equation}
Expanding $\twid{b}$ as $1 + \Gamma$ and ignoring terms higher than
quadratic in $\Gamma$ results in a contribution to (\ref{source}) of:
\begin{equation} \nonumber
\frac{e^{-\beta |p|}}{\sqrt{\alpha'}|p|} 
\bigl(\frac{8 \Gamma}{3} \bigr)
\, ,
\end{equation}

Since there are two such terms which yield identical contributions, the sum 
total of the contributions from the near massless states yields the equation of 
motion
\begin{equation}
\label{beomfinal}
\tddot{\Gamma} + 3H\tdot{\Gamma} + \frac{\mu}{M^3_5 2\pi a^3 |p|\alpha'^{3/2}} \bigl(\frac{8 \Gamma}{3} \bigr) = 0 \, ,
\end{equation}
where the exponential factor gets cancelled if we use this massless state as our 
reference state, as in (\ref{boltz}). The form of this equation clearly shows 
that $\Gamma$ will tend to zero if it starts out on either side of this value. 

However, to confirm that $\Gamma = 0$ is a genuine point of
equilibrium, we need to confirm that the extra massless states that appear at 
this radius (8 in all) contribute in such a way so that their sum vanishes. 
This can be verified by  a straightforward calculation 
\footnote{The only non-zero contribution to the driving term comes from the 
states $n=0, w=\pm 2, N=0$ and $n=\pm 2, w = 0, N = 0$ which make equal and 
opposite contributions and hence cancel.}. 

However, we wish to point out that as long as we stay in thermal equilibrium with 
the graviton gas, this equilibrium is actually metastable. The reason for this is 
easy to see from the formula for the mass of a winding mode:
\begin{equation}
\label{mass2}
\alpha'm^2 = \Bigl(\frac{n}{\twid{b}} + w\twid{b}\Bigr)^2 + 4(N-1) \, .
\end{equation}
In addition to the massless state given by $n=w=0$,$N=1$ 
(the graviton), and the 8 other massless states that appear at the self-dual 
radius (which are given by 
$N=1, n=-w=\pm1$; $N=0, w=n=\pm1$; $N=0,w=0,n=\pm2$ and $N=0,w=\pm2,n=0$), 
there are additional massless states at further special radii
\begin{eqnarray*}
\twid{b} = \frac{2}{|m|} &;&  w=\pm m,n=0,N=0\\\twid{b} = \frac{|m|}{2} &;& n=\pm m,w=0,N=0\\
m &\epsilon&  Z \, .
\end{eqnarray*}      
Thus, at half-integer multiples and and half integer fractions of the 
self-dual radius, two massless modes appear and will thus yield the dominant 
contributions to the driving term. These contributions again exactly cancel at 
twice the self dual radius, and at half the self dual radius. However in 
general, we will get a driving term that yields expansion at half integer 
points above twice the self dual radius and similarly, contraction at half 
integer fractions below half the self-dual radius. However since we posit that 
we begin at or near the self-dual radius, we are guaranteed to stay locked 
near it if our initial conditions satisfy
\begin{eqnarray*}
b(0) &\sim& \sqrt{\alpha'}\\
\tdot{b}(0) &\leq&  \sqrt{k} = \sqrt{\frac{2\mu_0}{M^3_5 a^3 2\pi \alpha'}} \, ,
\end{eqnarray*}
where the last condition constrains the initial ``velocity" of the scale 
factor to be such that it cannot roll over the ``hump" in the potential 
energy surrounding the metastable equilibrium at $b \sim \sqrt{\alpha'}$. 

Thus, we have demonstrated in the context of GR how a string gas in thermal 
equilibrium with a bath of gravitons and photons will \begin{it}dynamically 
stabilize the scale factor of the compact direction \end{it} if we begin 
close to that radius. Thermal equilibrium with the graviton bath ensures a 
persistent non-zero density of such winding modes. One can now imagine that, 
at some point, the winding mode gas becomes decoupled from the graviton gas, 
i.e. falls out of thermal equilibrium. In this situation, we are left with 
an unchanging driving term of the form (\ref{drive}), which yields the 
potential in Fig. 1, which will guarantee radion stabilization at the 
self dual radius for the remainder of the Universe's dynamics. We now turn our 
attention to the possible connection between  this
mechanism and inflationary and standard Big Bang cosmology.

\section{Late Time Evolution}

Recall the Einstein tensor for our metric setup:
\begin{eqnarray*}
G^0_0 &=& -3\hub{a} \Bigl[ \hub{a} + \hub{b} \Bigr] \\ G^i_j &=& -\delta^i_j \Bigl[ 2\hubd{a} + \hubd{b} + \Bigl( \hub{a} \Bigr)^2 +2\hub{b}\hub{a} \Bigr]\\ G^5_5 &=& -3 \Bigl[ \hubd{a} + \Bigl( \hub{a} \Bigr)^2 \Bigr] \, .
\end{eqnarray*}
We know that the dynamics of the scale factor $b$ in the situations we 
studied above cause it to undergo damped oscillations around the self dual 
radius. We demonstrated in a previous section how the ``spring constant" of 
this evolution will lock in to this equilibrium fairly rapidly. 
 
We can then study the evolution of the Universe after radius 
stabilization, which implies that $\tdot{b} = \tddot{b} = 0$ and 
$p-2r/3-\rho/3=0$. The resulting Einstein equations are:
\begin{eqnarray*}
G^0_0 = \frac{1}{M^3_5}T^0_0 &\to& H^2 = \frac{\rho}{3}\\
G^i_i = \frac{1}{M^3_5}T^i_i &\to& \tdot{H} = -\frac{1}{2}(\rho + p)\\
G^5_5 = \frac{1}{M^3_5}T^5_5 &\to& p - \frac{2r}{3} - \frac{\rho}{2} = 0\\
\nabla_\mu T^\mu_\nu = 0 &\to& \tdot{\rho} + 3H(\rho + p) = 0 \, ,
\end{eqnarray*}
where the $55$ equation is precisely the equilibrium condition. Thus, we 
recognize in the above the basic equations of FRW cosmology. We now consider 
how one achieves the two important epochs of late time FRW cosmology, namely 
the radiation dominated era and the matter dominated era.  

\subsection{Radiation Dominated Evolution}

If we assume that the density of 4-d matter gas is far greater than the 
density of the winding mode gas, then this will be the dominant component 
that drives the evolution of the macroscopic dimensions. 
If the 4-d matter has an equation of state parameter w, then the 
solutions to the Einstein equations become:
\begin{equation}
\label{einstmat}
\rho(t) \propto a^{-3(1+w)}
\end{equation} 
or
\begin{equation}
\label{einsta}
a(t) \propto t^{\frac{2}{3(1+w)}} \, .
\end{equation}
Thus, for a 4-d graviton and photon background, we get that 
$a(t) \propto t^{1/2}$, and so we reproduce a late time 
FRW evolution that is consistent with standard Big-Bang cosmology 
immediately after the end of inflation, whilst maintaining radius stabilization. 

\subsection{Matter Dominated Evolution}

Reconsidering (\ref{beom}):

\begin{equation}
\nonumber
\tddot{b} + 3H\tdot{b} + \frac{b}{M^3_{pl_5}}\Bigl(p - \frac{2r}{3} - \frac{\rho}{3} \Bigr) = 0 \, ,
\end{equation}

We see that any matter with the equation of state of non-relativistic 
dust ($p = 0$), can only drive the expansion of the radion if it is of a 
3-dimensional nature (i.e. $r \equiv 0$). This is surely to be a cause for 
concern when considering that at late times, one (naively) expects 3-dimensional 
non-relativistic dust to be the major driving component of the expansion of the 
universe, which would normally invalidate our stabilization mechanism in the 
present epoch.

However we wish to remind the reader that present day observations demand that a 
significant fraction of the energy density of the universe, which also drives the 
present day matter dominated expansion, be in the form of cold dark matter-- whose 
nature is as of yet completely unknown. There is a significant amount of interest 
is the prospect that extra-dimensional matter or extra dimensional effects might 
account for the `missing' matter in the universe. In what follows, we find that 
the only way to make our stabilization mechanism consistent with a 
`matter dominated' epoch is to introduce extra dimensional cold dark matter. 
We propose a candidate for this cold dark matter which is naturally contained in 
our framework, and discuss other possibilities which might have a natural 
realisation within the general brane gas framework (note that a similar proposal
was made in \cite{Gubser}). 

We see that in order for the driving term in (\ref{beom}) to correspond to a 
stable minimum at the self dual radius for matter which obeys the equation of 
state for non-relativistic dust ($p = 0$), we need to consider matter for which:
\begin{equation}
\label{cdm}
r = -\rho/2|_{b = \sqrt{\alpha'}} \, .
\end{equation}
That is, we require the dominant component of the energy density which is 
driving the expansion of the universe be such that it preserves the stability 
of the radion at the self dual radius. Matter which exhibits such an equation 
of state will surely have to be massive (else there will be a non-zero pressure 
along the non-compact directions for any non-zero energy density). In addition, 
such matter will have to be something beyond presently supposed dark matter 
candidates (WIMPS, supersymmetric relics etc.) as it will neccesarily have to be 
`extra-dimensional' in nature. We now show that our model naturally contains such 
a candidate. Recalling the discussion surrounding (\ref{bdd}), we see that the 
equation of state parameter for a gas of strings with a particular set of quantum 
numbers is obtained from the following equation:
\begin{equation}
\label{eoscdm}
r = \rho \times \frac{ \frac{n^2}{b^2} - \frac{w^2 b^2}{\alpha'^2}}{p^i p_i 
+ \frac{4}{\alpha'}(N - 1) + (\frac{n}{b} + \frac{wb}{\alpha'})^2} \, ,
\end{equation}
where the momentum will be set to zero (or is vanishingly small) in order to 
satisfy $p = 0$ (c.f. (\ref{p2})). In particular, since we are looking for 
states which can satisfy (\ref{cdm}) at the self dual radius, we need to find 
the appropriate quantum numbers which have an equation of state parameter 
$w = -1/2$ at $b = \sqrt{\alpha'}$, which reduces to the following condition:
\begin{equation}
\label{cdmqn}
3n^2 - w^2 + 4(N-1) + 2nw = 0 \, ,
\end{equation}
and we have to be mindful of the level matching constraint: $N + nw \geq 0$. 
As expected, it turns out that the massless states that satisfy these conditions 
have an energy density proportional to $|p|$, whereas the pressure is proportional 
to $|p|^2/3$, and hence one cannot have a non-zero pressure without having a 
vanishing energy density. 

The first massive states which satisfy (\ref{cdmqn}) are 
represented by the quantum numbers:
\begin{equation}
\label{qns}
|p| = 0~,~N = 2~;~n = 0 ~,~ w = \pm 2 \, .
\end{equation}
These states contribute to the energy momentum tensor as follows 
(c.f. (\ref{rho2}) - (\ref{r2})):
\begin{equation}
\label{massive}
p = 0~,~ \rho = \frac{2\mu\sqrt 2}{2\pi a^3\alpha'}~,~ r = -\rho/2 \, ,
\end{equation}
where the factor of $\alpha'$ in the denominator comes from two factors of 
$\sqrt{\alpha'}$-- one from the metric factor $b$ which has stabilized at the 
self dual radius, and the other as the pre-factor of the non zero rest mass of 
this string state. As we will see in the next section, were we to look at 
fluctuations of the radion around the self dual radius, these states also 
provide a stable equilibrium at $b = \sqrt{\alpha'}$ in a phenomenologically 
acceptable manner. Thus, taking questions of consistency with observation 
for granted for the moment, we see if these particular string states are taken to 
dominate the present energy density of the universe, then by the Einstein 
equations derived at the start of this section, we can admit an epoch of dust 
driven FRW expansion ($p = 0$, $a \propto t^{2/3}$) at late times, consistent 
with radius stabilization. 

   However, there are many further issues that will have to be resolved if we 
are to take this idea of stringy dark matter seriously, which we postpone to a 
future report. At present, however, we wish to state that there are indications 
that such stringy dark matter might have the right clustering properties at the 
level of first order perturbation theory, in that local overdensities of this 
stringy dark matter induces gravitational clustering in the remaining 3-dimensions.

We also wish to point out that in certain situations, non-relativistic 
p-branes might also be able to provide us with a matter content with 
satisfies (\ref{cdm}) \cite{Boehm}. In the context of brane gas cosmology, this 
is an appealing idea as one might need higher dimensional branes to stabilize 
compact sub-manifolds that do not admit topological one-cycles (and hence do 
not admit winding modes). We will investigate this possibility further in a 
future report. Finally, we wish to address the effects of an intermediary 
epoch of scalar field driven inflation on our stabilization mechanism.

\subsection{Intermediate (Non-Stringy) Inflation} 

We find that our mechanism for radius stabilization might not be 
compatible with an intermediate epoch of bulk scalar field driven inflation. 
To investigate this, we first consider the energy-momentum tensor of an 
almost homogeneous inflaton field:
\begin{equation}
\label{inflatonem}
T^\mu_\nu = diag \Bigl(-\bigl[\frac{\tdot{\phi}^2}{2}+V(\phi) \bigr],\bigl[\frac{\tdot{\phi^2}}{2}-V(\phi) \bigr],...,\bigl[\frac{\tdot{\phi^2}}{2}-V(\phi) \bigr] \Bigr)  
\end{equation}
Adding this energy-momentum tensor to our string gas yields a 
non-trivial contribution to the driving term in the equation of motion
for $b$. This driving term then takes the form
\begin{eqnarray}
\label{inflatondrive}
&-&\frac{2b}{3M^3_5}V(\phi) + \frac{\mu_{ref} e^{\beta E_{ref}}}{M^3_{pl_5} a^3 2\pi} \\
&& \sum_{n,w,N,p^2} \frac{e^{-\beta E}}{\sqrt{E}} \Bigl( -\frac{n^2}{b^2} - \frac{2nw}{3\alpha'} + \frac{w^2b^2}{3\alpha'^2} - \frac{4(N-1)}{3\alpha'} \Bigr)
\nonumber
\end{eqnarray}  
from which it is easy to see that the inflaton contribution will drive 
expansion in the extra dimension. In general, this term will compete 
with the string gas contribution which, as we have seen, drives contraction, 
if we are above the self-dual radius. However, this competition is short 
lived, as the factor of $a^3$ in the denominator of the string gas driving 
term will quickly render it irrelevant and the scale factor will then 
expand according to
\begin{equation}
\label{inflbeom}
\tddot{b} + 3H\tdot{b} -\frac{2b}{3M^3_5}V(\phi) = 0 \, .
\end{equation} 
Recalling that during this (slow roll) inflation $H$, and by the time-time 
Einstein equation, also $V(\phi)$ are almost constant, we can solve the above 
equation, with the resulting two solutions:
\begin{eqnarray*}
b(t) &\propto& exp^{-\frac{3H}{2} \bigl( 1 +\sqrt{1+\frac{8V(\phi)}{9H^2 3M^3_5}}\bigr)t}\\
b(t) &\propto& exp^{-\frac{3H}{2} \bigl( 1 -\sqrt{1+\frac{8V(\phi)}{9H^2 3M^3_5}}\bigr)t} \, .
\end{eqnarray*}
Substituting in the Einstein equation $H^2 = V(\phi)/3M^3_5$ gives us
\begin{eqnarray*}
b(t) &\propto& exp^{-\frac{3H}{2} \bigl( 1 +\sqrt{1+\frac{8}{9}}\bigr)t} \propto e^{-3.56Ht}\\
b(t) &\propto& exp^{-\frac{3H}{2} \bigl( 1 -\sqrt{1+\frac{8}{9}}\bigr)t} \propto e^{0.56Ht} \, .
\end{eqnarray*}
Except for very special initial conditions, the growing mode will
rapidly come to dominate. Thus, we conclude that $b$ expands 
exponentially (though not as fast as $a$). 
After inflation has finished (and after reheating to a temperature
smaller than the one required for the pair production of string winding
modes), $a$ expands as $t^{1/2}$. The energy density in the string gas 
will have been exponentially
suppressed by the inflationary evolution, and thus to good approximation
the equation of motion for $b$ will take the form  
\begin{equation} \nonumber
\tddot{b} + 3H\tdot{b} = 0 
\end{equation}
leading to
\begin{equation}
\tdot{b} = Ca^{-3}
\end{equation}
which implies
\begin{equation}
b(t) = b(1) + 2\tdot{b}(1) \Bigl( 1 - \frac{1}{\sqrt{t}} \Bigr) \, , 
\end{equation}
where $(1 \leq t)$. Thus, $b$ asymptotically expands to some limiting, and 
very large (due to the initial conditions that result at the end of 
inflation) value. 

In conclusion, we have seen how our radion stabilization mechanism is 
consistent with the FRW evolution of the non-compact dimensions, but not 
with an intermediate inflationary period, with inflation driven by
a bulk scalar matter field. Thus, in order for brane gas cosmology to make
contact with the present cosmological observations, one either needs
a different (maybe stringy) realization of inflation (see \cite{BEM}
for some ideas) where strings are produced in re-heating, or a non-inflationary 
mechanism to solve the 
cosmological problems of standard Big Bang cosmology and to produce
a spectrum of almost scale-invariant cosmological perturbations. Finally, we turn 
to various phenomenological issued pertaining to this model.

\section{phenomenology} 

There are two potential phases of applicability of our considerations. The first
is to the early phase of string gas cosmology before a period of inflation. In
this case, there are no phenomenological constraints on the model since the
number density of the particles (from the four dimensional perspective) 
which correspond to the string states wrapping the extra dimension are diluted
exponentially during inflation. However, in this case our considerations would
no longer be relevant for the late-time stabilization of the extra dimension.

The second phase of potential applicability of our considerations is to the
universe after inflation of our three spatial dimensions. We then need to 
assume that winding and momentum modes about the extra dimension can be
regenerated in sufficient number, as discussed in \cite{Gubser}. In this case,
there are important constraints on our model. We must ensure that the particles
corresponding to our string states do not overclose the universe. In addition,
there is a radion mass constraint. Since the radion corresponds to a scalar
particle from our four-dimensional perspective, we must make sure that its mass
is consistent with the experimental constraints (we thank the Referee for
stressing this point to us).
 
An additional constraint comes from the string theoretical aspect of our model: 
we must ensure that the derivatives of the metric remain several orders of 
magnitude smaller than the worldsheet derivatives (see Appendix). This is to 
ensure that we can inherit aspects of the string spectrum and constraint algebra 
that we have used so far. From (\ref{beomfinal}), which has the form:
\begin{equation}
\label{form}
\ddot{\Gamma} + 3H\dot{\Gamma} + k\Gamma = 0
\end{equation}
we see that the `spring constant' which sets the scale for the how fast the 
metric factor $b$ varies, is given by
\begin{equation}
\label{springk}
k = \frac{8\mu_0}{3M_5^3 \alpha'^{3/2}|p|a^3} \,
\end{equation}
where the subscript on $\mu$ is to remind the reader that any explicit 
metric dependence has already been factored out (see section II). We
have taken the stabilization to be provided by the massless states discussed 
earlier with $|p|$ denoting the momentum in the non compact directions, 
$N = 1$, and $n = -w = \pm1$. 
In order that our metric factors evolve much slower than the string scale, we 
require that $\partial_\mu g \ll 1/\sqrt{2\pi\alpha'}$. Since (\ref{beomfinal}) is 
a second order ODE, this implies that $k \ll 1/2\pi\alpha'$. As discussed in the 
Appendix, we choose to be quite conservative and demand that 
$k \leq 10^{-6}/2\pi\alpha'$. 

A second constraint comes from requiring that the stabilization mechanism be 
effective at all times. This leads to a lower bound on $k$. We take this bound
to be given by the `critically damped' value for $k$:
\begin{equation}
k_{crit} \, = \, 9H^2 \, .
\end{equation}  

The two above constraints yield the following bounds:
\begin{equation}
\label{phen1}
9H^2 \leq \frac{8\mu_0}{3M_5^3 \alpha'^{3/2}|p|a^3} \leq\frac{10^{-6}}{2\pi\alpha'}
\, .
\end{equation}

Since the winding states that stabilize the extra dimensions are massless at the
self-dual radius, they will behave as hot dark matter --  dark matter because
they only interact gravitationally (through the tree level interactions 
$w +\stackrel{\_}{w} \to h_{\mu \nu}$) with other fields, hot because they
are massless and have a radiative equation of state. We have to ensure that we do 
not introduce too many of these objects so that we can ensure consistency with 
observational bounds. 

Next, we have to ensure that the massive string states that we propose as 
a candidate for the cold dark matter that is presently driving the `dust dominated' 
expansion of the universe do not violate any observational bounds while preserving 
our stabilization mechanism.   

Let us begin by considering the massless states which are presently stabilizing 
the radion. From (\ref{end}) we see that post stabilization, for stringy matter
with $N = 1$, $n = -w = \pm1$, we must have:
\begin{equation}   
\label{criti}
\rho = \frac{\mu_{0} |p|}{a^3} \leq 10^{-4}\rho_{crit} \, 
\end{equation}
in order to ensure consistency with the nucleosynthesis bounds.
The critical density of the universe is $\rho_{crit} = 10^{-29} g/cm^3$. 
We then find that (\ref{criti}) becomes
\begin{equation}
\label{criti2}
\mu_0  \, \lesssim \, 10^{-4} 10^{-37} {\rm GeV}^4 |p|^{-1} \, .
\end{equation} 
Let us now parametrize the present value of $|p|$ as
\begin{equation}
|p| \, = \, 10^{-\gamma} {\rm eV} \,
\end{equation}
where $\gamma$ is some constant determined by the initial conditions and
the history of the universe, and
this parametrization being motivated by the fact that 
$|p|$ is likely to be of the order of a few eV in the present 
epoch if it corresponds to an initial $|p|$ of the order of the Planck 
energy. Then, the above bound takes the form
\begin{equation}
\label{phen2}
\mu_0 \, \lesssim \, 10^{-41}{\rm GeV}^3 10^{\gamma}.
\end{equation}

On the other hand, the first inequality in (\ref{phen1}) becomes:
\begin{equation}
\label{finalbound}
\mu_0 \, \geq \, H^2(t_0) |p| \, \sim \, 10^{-93} 10^{- \gamma} {\rm GeV}^3 \,
\end{equation}
which is consistent with (\ref{phen2}). In the above, we have made use of
\begin{equation} \label{help}
M_5^{-3} \, = \, 8 \pi G_5 \, = \, 8 \pi G_4 \sqrt{\alpha'} \, .
\end{equation}
Using (\ref{help}), it can easily be checked that the upper bound on $\mu_0$
which follows from the second inequality in (\ref{phen1}) is much weaker than
the bound (\ref{phen2}). 
We conclude that one can easily arrange the number
of string modes such that stabilization of the extra dimension is ensured
and at the same time the massless modes do not conflict with the nucleosynthesis
bound.

Let us next turn to the radion mass constraint: Since the radion appears in
four dimensions as a scalar field, its mass must be larger than
\begin{equation}
m_{crit} \, = \, 10^{-12} {\rm GeV}
\end{equation}
in order to avoid fifth force type constraints. Since the square of the
radion mass is given by $k$, this constraint becomes
\begin{equation} \label{mubound}
\mu_0 \, \geq \, M_5^3 \alpha'^{3/2} |p| m_{crit}^2 \, 
\sim \, 10^{-35 - \gamma} GeV^{3}
\end{equation}
which is consistent with the upper bound (\ref{phen2}) on $\mu_0$ if
$\gamma \geq 3$. Such a value of $\gamma$ is not at all unreasonable and
could easily arise from an additional suppression of the momentum during 
a short period of inflation.

It turns out to be crucial that we use massless modes to stabilize the extra 
dimensions, as more massive string states would bring down the upper bound, 
and as we are about to see, do not provide as effective a spring constant and 
hence bring up the lower bound, to the net effect that it is phenomenologically 
inconsistent to have these as the only strings that are stabilizing the radion. 
We arrive at this observation by considering the second aspect of our 
phenomenology-- namely, that we would like the cold dark matter content of our 
present universe to consist of massive string modes (which satisfy (\ref{cdm}), 
which as we have seen is require in order to maintain stabilization at the self 
dual radius). 

Considering the contribution to the energy density by a string gas with quantum 
numbers $|p|=0$, $N=2$, $n=0$, $w=\pm 2$ (\ref{massive}), and equating this to 
the critical density of the universe, we see that:
\begin{equation}
\label{dmcrit}
\frac{2\sqrt{2}\mu_{dm}}{2\pi a^3 \alpha'^{1/2}} \, \approx \, \rho_{crit} 
\end{equation}     
Where now the subscript on $\mu$ serves to indicate that this is our dark matter 
candidate. This requires
\begin{equation}
\mu_{dm} \, \sim \, 10^{-67} {\rm GeV}^3 \, .
\end{equation}
Upon perturbing around a stabilized radius, we find that these string 
modes contribute to the stability of the radion with the spring constant:
\begin{equation}
\label{scdm}
k_{dm} = \frac{8\pi G_{5} \mu_{dm}}{2\pi \alpha' a^3}\frac{\sqrt 8}{3} \, .
\end{equation}
Demanding that this value of $k$ is consistent with the radion mass bound
yield a lower bound on $\mu_{dm}$ 
\begin{equation}
\mu_{dm} \, \geq \, m_{crit}^2 M_4 \, \sim \, 10^{-6} {\rm GeV}
\end{equation}
which is clearly inconsistent with (\ref{dmcrit}) for values of $\gamma$ which
are extremely large. 

Thus we see that if we 
introduce the correct amount of our dark matter candidate, it contributes too 
feebly to the dynamics of the radion (even though it does provide its own 
contribution to stabilization). However, this is of no concern to us, as we have
already shown that the massless string states provide a robust stabilization 
mechanism that is consistent with observational bounds. Thus, if our string 
gas has a massive component that serves as cold dark matter, and a massless 
component that stabilizes the radion (and behaves like hot dark matter) in the 
right proportions, which as we have shown is quite easy to achieve, we can be 
assured of the phenomenological consistency of our stabilization mechanism with 
late time FRW cosmology.

\section{Conclusions}

The analysis in this paper is motivated by brane gas cosmology \cite{BV,ABE}.
As a simplified problem, we have studied the effects of a gas of strings with
non-vanishing momentum and winding modes about a single compact 
extra dimension taken to be a circle (the three large dimensions
are taken to be spatially flat and isotropic) on the evolution of 
the radius of that dimension, 
assuming that the background space-time satisfies the equations of motion 
of General Relativity. We discovered that such a string gas
leads to a dynamical stabilization mechanism for the radius of this
dimension, the radion. Assuming initial conditions in which the three
large dimensions are expanding, we found that the radion performs
damped oscillations about the self-dual radius. 

In a first step, we studied the effects of a gas of non-interacting
strings, each string having the identical momentum and winding numbers.
Key to the stabilization mechanism is the fact that winding modes
and momentum modes contribute with opposite sign to the pressure of the
string gas in the direction of the compact dimension, and that the
winding modes generate a potential for the radion which favors
contraction, whereas the momentum modes generate a potential favoring
expansion. We then showed that the stabilization mechanism also holds
for a gas of strings in thermal equilibrium. 

We also showed that, after radion stabilization, the scale factor
for the three large spatial dimensions obeys the usual FRW equations
of standard Big Bang cosmology. Thus, our scenario leads in a natural
way to a late time FRW Universe. However, we have also shown that
the radion stabilization mechanism is not compatible with a period
of scalar field driven bulk inflation. Thus, in order for brane gas
cosmology to make successful contact with present cosmological
observations, one either needs to find a stringy mechanism for driving
inflation where strings are produced in the re-heating epoch, or else one must 
provide an alternative to inflationary cosmology, both for solving the mysteries 
of standard Big Bang cosmology,
and for explaining the origin of the observed large-scale fluctuations.
 
Note that we start with the assumption that three spatial dimensions
are already much larger than the other ones (one other dimension in
our case). Whether or not the dynamics of strings in the initial
stages will indeed lead to this situation may depend on the corner
of M-theory one is working in, i.e. on the specific form of the
background equations of motion and initial conditions 
(see \cite{BV,ABE,BEK,Col2,Col3} for different angles on this issue).
However, it should not be hard to generalize our analysis to a
situation with more compact dimensions or different topologies
and geometries of the extra dimensions, which will be the focus of our future 
work. 

In future work we also plan to study the annihilation rate of the string modes
which are central to this work, namely modes which have both winding and momentum
in the compact direction. Since these modes interact only gravitationally
just like gravitons, they will be out of thermal equilibrium at late times
and hence will not decay.

\vskip.5cm
\centerline{\bf Acknowledgements}

We wish to thank Thorsten Battefeld, Tirthabir Biswas and 
Scott Watson for many useful 
discussions and criticisms, and for important comments on the
manuscript. SP wishes to thank Nikolaos Daniilidis help 
rendered at various points in this study. This research is partly 
supported by the U.S. Dept. of Energy under Contract DE-FG0291ER40688, 
Task A.

\begin{appendix}

\section{The String Spectrum in a Time Dependent Background}

Let us then begin with the Polyakov action for a closed string:
\begin{equation}
\label{polyakov}
S = \frac{-1}{4\pi \alpha'} \int d^2 \sigma \sqrt{-\gamma} \gamma^{ab} \partial_a X^\mu \partial_b X^\nu g_{\mu \nu}(X) \, .
\end{equation}
Varying this action with respect to the world sheet metric gives us the 
equation of motion
\begin{equation}
\label{polyeom1}
\gamma_{ab} = \lambda h_{ab} = \lambda \partial_a X^\mu \partial_b X^\nu g_{\mu \nu}(X) \, ,
\end{equation}
where we can exploit the world-sheet diffeomorphism and Weyl invariance to 
make the world-sheet metric flat (conformal gauge):
\begin{equation}
\label{polyeom2}
\gamma_{ab} = \gamma^{ab} = \begin{pmatrix} &-1 & 0 \cr &0 & 1\end{pmatrix} \, .
\end{equation}
Varying the action with respect to the world-sheet fields, and imposing the 
gauge choice yields the equations
\begin{equation}
\label{polyeom3}
\partial_a ( \partial^a X^\lambda g_{\lambda \mu}(X) ) = \frac{1}{2} \partial^a X^\lambda \partial_a X^\nu g_{\lambda \nu,\mu}(X) \, ,
\end{equation}
where the meaning of the derivatives of the metric should be clear. 
This equation translates into
\begin{equation}
\label{polyeom4}
\partial_a \partial^a X^\tau + \Gamma^\tau_{\lambda \nu} \partial_a X^\nu \partial^a X^\lambda = 0 \, .
\end{equation}

Now, we consider the case when our metric depends only on time, that is 
$g_{\mu \nu}(X) = g_{\mu \nu}(X^0)$, is diagonal and has -1 as its $00$ entry
(This last point is not essential to the argument, it only serves to 
simplify the equations). Unpacking the above equation yields the equations
\begin{equation}
\label{polyeom5}
-\partial_\tau^2 X^0 + \partial_\sigma^2 X^0 = -\frac{1}{2} g_{\lambda \nu,0}( -\partial_\tau X^\nu \partial_\tau X^\lambda + \partial_\sigma X^\nu \partial_\sigma X^\lambda ) \, ,
\end{equation}
and
\begin{equation}
\label{polyeom6}
-\partial_\tau^2 X^i + \partial_\sigma^2 X^i = -g^{ii}g_{ii,0}( -\partial_\tau X^i \partial_\tau X^0 + \partial_\sigma X^i \partial_\sigma X^0 ) \, .
\end{equation}
On the right hand sides of the equations there is an overall factor
containing the time derivative of the metric.
To estimate the magnitude of either side of the 
equations, we realize that world-sheet time derivatives will be of the 
order of the energy of the string, which is of the order of the 
square root of the string 
tension: $\partial_\tau \sim \frac{1}{\sqrt{\alpha'}}$. 
Similarly, the world-sheet 
spatial derivative will be of the order of the inverse of the string 
length $l_s$, $\partial_\sigma \sim \frac{1}{l_s}$, which is of the
order of the square root 
of the string tension. On the right hand sides, the terms are 
of the same order as the terms on the left, except for the factors containing
derivatives of the metric. As long as these metric derivatives are several 
orders of magnitude smaller than the string tension scale, we can safely 
ignore them \footnote{For instance, we are tempted to be conservative and to 
ask for them to be roughly 6-8 orders of magnitude smaller, in order to be 
certain of a consistent treatment.}. Assuming such a background, 
(\ref{polyeom5}) and (\ref{polyeom6}) reduce to the usual flat space wave 
equations, and we can proceed to expand the solutions in terms of plane 
waves. We then impose the canonical commutation relations on the expansion 
coefficients \footnote{Recall that $\partial_\tau X^\mu:=P_\mu$ is the 
canonical momentum, where we must view $P_\mu$ as having its 
index lowered which is to be raised with the inverse metric. Not realizing 
this will produce nonsensical results elsewhere (such as in the computation 
of the energy-momentum tensor of the string) in addition to making 
quantization very awkward. This fact is easier to understand if we recall 
that $P_\mu$ and $X^\mu$ are canonically conjugate world sheet fields and 
not 4-vectors. With 
this in mind, canonical quantization means imposing 
$[P_\mu(\tau,\sigma),X^\nu(\tau,\sigma')] = -i \delta^\nu_\mu\delta(\sigma - 
\sigma')$ which, as is, does not involve the metric. Hence we do not have 
to do anything different at this stage of the analysis.}. 

To complete the 
analysis, we impose the constraints coming from our gauge choice:
\begin{equation}
\label{gaugechoice}
h_{ab} = \begin{pmatrix} -1 & 0 \cr 0 & 1\end{pmatrix}
\end{equation}
\begin{eqnarray} \label{constraint1}
\partial_\tau X^\mu \partial_\tau X^\nu g_{\mu \nu}(X) 
+ \partial_\sigma X^\mu \partial_\sigma X^\nu g_{\mu \nu}(X) \, 
&=& \, 0 \, ,  \\ 
\partial_\tau X^\mu \partial_\sigma X^\nu g_{\mu \nu} (X) \, &=& \, 0 \, .
\end{eqnarray}  
By implementing our plane wave expansion for $X^\mu$ in terms of the 
creation and annihilation operators, we obtain the spectrum by requiring 
that physical states are annihilated by the constraint operators. However, 
when we write the constraints in terms of the canonical variables, their 
metric dependence becomes manifest:
\begin{equation}
\label{virasoro}
P_\mu X'^\mu = 0 
\end{equation}
\begin{equation}
\label{L0}
P_\mu P^\mu + X'_\mu X'^\mu = 0 \, .
\end{equation}   
We see that (\ref{virasoro}) is independent of the metric. Since this 
constraint is the origin of the Virasoro algebra, we see that it remains 
valid in suitably time-dependent backgrounds. However (\ref{L0}) does 
depend on the metric since $P^\mu$ requires the inverse metric to raise its 
index and $X'_\mu$ requires the metric to lower its index. Now we choose to 
work in a 5-d background with metric:
\begin{equation}
\label{metric}
g_{\mu \nu} = diag(-1,a^2(t),a^2(t),a^2(t),b^2(t)) \, ,
\end{equation}     
where the 5'th dimension is taken to be compact with radius 
$2\pi b$ \footnote{The results in our paper easily generalize to backgrounds 
of any number of non-compact dimensions so long as precisely one dimension 
is compactified on a circle. For consistency, we could state that our 
background is a compactified 10 dimensional space with six dimensions 
compactified on a Calabi-Yau (CY)space and one dimension compactified on a 
circle. Since in the prototypical compactification scenarios such as
the Horava-Witten model \cite{HW} the radius of the CY is smaller than
the radius of the circle, since we will be interested in string winding 
modes but Calabi-Yau spaces do not admit one-cycles, we can ignore the presence
of the Calabi-Yau space if we work 
in an effective Lagrangian description valid at energies smaller than
the energy scale of CY compactification. We could always go back and 
work in a $9+1\times S^1$ or $24+1\times S^1$ space-time where we would 
derive the same conclusions as we do here.}. With this as our background 
metric, (\ref{L0}) becomes (for a string wound along the 5'th dimension):
\begin{eqnarray}
\label{energy0}
-E^2 + g^{ij}p_i p_j &+& \frac{2}{\alpha'}(N + \stackrel{\_}{N} - 2) \\
&+& g^{55}P_5 P_5 + g_{55}X'^5 X'^5 = 0 \, ,\nonumber
\end{eqnarray}
where all we have done is expanded out (\ref{L0}), and realized that the 
terms coming from the non-compact $X^\mu$ and $P_\mu$ give us the center
of mass momenta and the left and right oscillator pieces, and the terms 
coming from $X^5$ have been accounted for explicitly. We know that this 
part of the energy contributes \cite{Pol}: 
\begin{equation}
\label{winding}
P_5P^5 + X'_5X'^5 = \frac{n^2}{b^2} + \frac{w^2 b^2}{\alpha'^2} \, ,
\end{equation}
so that combined with the level matching conditions
\begin{equation}
\label{level1}
nw + N - \stackrel{\_}{N} = 0 
\end{equation}
we get
\begin{equation}
\label{energy1}
E = \sqrt{ g^{ij}p_i p_j + \frac{4}{\alpha'}(N - 1) + (\frac{n}{b} + \frac{wb}{\alpha'})^2} \, ,
\end{equation} 
where the only remnant of the level matching condition is the requirement that
\begin{equation}
\label{level2}
N + nw \geq 0 \, .
\end{equation}

Thus, we wee that the only effect of working in a slowly varying background 
is to introduce time-dependent metric factors in the obvious places in 
(\ref{energy1}) which is otherwise identical to the result we would obtain 
in a static background. 

\end{appendix}

\end{document}